\documentstyle[twoside,fleqn,espcrc2]{article}

\def\lb{\lbrack}
\def\rb{\rbrack}

 \def\Slash#1{
  \begin{picture}(5,6)(0,0)
  \put(-.7,-1.2){\line(5,6)6}
  \end{picture}
  \kern-.8em#1}
 \def\slash#1{
  \begin{picture}(5,6)(0,0)
  \put(-1.5,-1.7){\line(5,6)5}
  \end{picture}
  \kern-.8em#1}

\def\sd{\Slash \partial}

\def\Tr{\mbox{Tr}}

\def\g5{\gamma_5}
\def\hg5{\hat{\gamma}_5}
\def\A{{\cal A}}
\def\index{\mbox{index}\,}

\def\C{{\cal C}}

\def\G{{\cal G}}
\def\U{{\cal U}}

\def\hC{\widehat{\cal C}}

\def\ch{\mbox{ch}}

\def\Qlatmr1{Q_{lat}^{(m=r=1)}}

\def\be{\begin{eqnarray}}
\def\ee{\end{eqnarray}}

\def\SU{\mbox{SU}}
\def\index{\mbox{index}}
\def\ch{\mbox{ch}}

\newcommand{\AmS}{{\protect\the\textfont2
  A\kern-.1667em\lower.5ex\hbox{M}\kern-.125emS}}

\title{Fermionic topological charge of families of lattice gauge fields}

\author{David H. Adams\address[MCSD]{Physics dept., Box 90305, Duke University, \\ 
        Durham, NC 27708, USA}%
        \thanks{Current address: Instituut Lorentz, Universiteit Leiden, P.O. Box 9506,
                NL-2300 RA Leiden, The Netherlands}}
       
\begin{document}

\begin{abstract}
Topological charge of families of lattice gauge fields is defined fermionically via families index theory
for the overlap Dirac operator. Certain obstructions to gauge invariance of the overlap chiral fermion 
determinant, as well as the lattice analogues of certain obstructions to gauge fixings without the Gribov
problem, have natural descriptions in this context.
\vspace{1pc}
\end{abstract}

\maketitle

\section{Continuum setting}

Continuum SU($N$) gauge fields on ${\bf R}^4$ with finite YM action are pure gauge at infinity:
$A_{\mu}(x)\to\phi(x)\partial_{\mu}\phi(x)^{-1}$ for $|x|\to\infty$. Such fields have a well-defined
topological charge 
\be
Q:=\mbox{deg}(\phi)\,,
\label{1}
\ee
the degree of $\phi:S^3\to\SU(N)$, where $S^3$ is the 3-sphere ``at infinity'' in ${\bf R}^4$.
The topological charge has a fermionic description: by the Index Theorem, $Q=\index(\sd^A)$ 
where $\sd^A$ is the Dirac operator coupled to $A$. The situation is similar for 
gauge fields on compact manifolds such as $S^4$, $T^4$, $\dots$

Consider now a family $A^{(y)}$ of $\SU(N)$ gauge fields on the 4-torus $T^4$ parameterized by 
$y\in{}Y$ (a smooth parameter space) with the property that the gauge fields parameterized by boundary
points of $Y$ are all gauge equivalent:
\be
A^{(y)}=\phi^{(y)}\cdot{}A\qquad\mbox{for}\ y\in\partial{}Y
\label{1a}
\ee
where $\phi^{(y)}$ is a family of gauge transformations parameterized by $\partial{}Y$. 
Then the family $A^{(y)}$ determines a closed submanifold in the orbit space
of $\SU(N)$ gauge fields on $T^4$ and there is a natural notion of topological charge for the family:
\be
Q_Y:=\mbox{deg}(\Phi)
\label{2}
\ee
where $\Phi:\partial{}Y\times{}T^4\to\SU(N)$ is given by $\Phi(y,x)\!:=\!\phi^{(y)}(x)$. This families
topological charge also has a fermionic description: by the Families Index Theorem of \cite{AS},
\be
Q_Y=\int_{Y/\partial{}Y}\,\mbox{ch}(\index\,\sd)\,.
\label{3}
\ee
Here $Y/\partial{}Y$ denotes $Y$ with its boundary collapsed to a single point and 
$\mbox{ch}(\index\,\sd)$ is the Chern character of the index bundle of $\sd$ over the gauge orbit space.

We henceforth specialize to the case where $Y$ is a $2p$-dimensional ball 
$B^{2p}\,$, $\partial{}Y=S^{2p-1}\,$, then the family $A^{(y)}$ corresponds to a $2p$-sphere in
the orbit space. We denote the families topological charge in this case by $Q_{2p}$. Examples of such
families can be constructed as follows. Choose a map $\Phi:S^{2p-1}\times{}T^4\to\SU(N)$ and some gauge 
field $A$, then set
\be
A^{(y)}\;\equiv\;A^{(\theta,t)}:=t(\,\phi_{\theta}\cdot{}A)
\label{4}
\ee
where $\theta\in{}S^{2p-1}\,$, $t\in[0,1]$ is the radial coordinate in $B^{2p}$, and 
$\phi_{\theta}(x)\!:=\!\Phi(\theta,x)$. Then $Q_{2p}\!=\!\mbox{deg}(\Phi)$ independent of the choice of $A$.
Maps $\Phi$ with nonvanishing degree exist for $1\!\le{}\!p\!\le{}\!N\!-\!2$.

{\it Physical significance of the $Q_{2p}$'s}. The physical significance of the usual $Q\!=\!Q_0$ is
well-known so we concentrate on the other cases. $Q_2$ is an obstruction to gauge invariance of the
chiral fermion determinant \cite{AG,AS}; this follows from the fact that the degree 2 part of
$\ch(\index\,\sd)$ coincides with the Chern character of the determinant line bundle of $\sd$. 
The general $Q_{2p}$'s for $p>0$ also have significance as obstructions to the existence of gauge 
fixings without the Gribov ambiguity \cite{DA(NPB2)}: Such a gauge fixing picks out a submanifold
$\A_f$ of the space $\A$ of gauge fields which intersects each gauge orbit precisely once, thus determining
a decomposition
\be
\A\;\simeq\;\A_f\times\G\;\simeq\;\A/\G\times\G
\label{5}
\ee
Since $\A$ is an affine vector space all spheres in $\A$ are contractible; the decomposition (\ref{5})
then implies that all spheres in $\A/\G$ and $\G$ are contractible. In this case the families topological
charge $Q_{2p}\,$, which via (\ref{3}) can be expressed as 
\be
Q_{2p}=\int_{{\cal S}^{2p}}\,\ch(\index\,\sd)
\label{6}
\ee
where ${\cal S}^{2p}$ is the $2p$-sphere in $\A/\G$ determined by the $B^{2p}$-family $A^{(\theta,t)}\,$,
vanishes, since, being an integer, it is unchanged under a smooth contraction of ${\cal S}^{2p}$
to a point in $\A/\G$. Thus, nonvanishing $Q_{2p}$ ($p>0$) implies nonexistence of a decomposition
(\ref{5}), i.e. nonexistence of a gauge fixing without the Gribov problem.

{\it Remark}. The decomposition (\ref{5}) means that for each $A\in\A$ there is a unique
$A_f\in\A_f$ and $\phi\in\G$ such that $A=\phi\cdot{}A_f$. For $\phi$ to be unique $\G$ must act
{\em freely} on $\A$. To achieve this we restrict $\G$ to be the group of gauge transformations
satisfying $\phi(x_0)\!=\!1$ for some arbitrarily chosen basepoint $x_0$ in $T^4$. This has the effect of 
excluding the nontrivial constant gauge transformations. (The same condition was imposed in  
the families index theory for the Dirac operator in \cite{AS}.) Thus our setting is
different from the one in Singer's study of the Gribov problem in \cite{S}. There $\A$ was taken
to consist of the {\em irreducible} gauge fields, which are acted freely upon by $\G/{\bf Z}_N\,$
(${\bf Z}_N$=the center of $\SU(N)$) and the obstructions in that setting are different from the
$Q_{2p}$'s discussed above. (The obstruction studied by Singer was $\pi_1(\G/{\bf Z}_N)$.)

\section{Lattice setting}

To begin with the space of lattice gauge fields is
\be
\U_{\rm initial}\,\simeq\,\SU(N)\times\SU(N)\times\cdots\times\SU(N)
\label{6a}
\ee
(one copy for each lattice link). Topological charge for lattice gauge fields can be defined 
fermionically as
\be
Q:=\index\,D^U
\label{7}
\ee
where $D^U$ is the overlap Dirac operator \cite{Neu} coupled to the lattice gauge field $U$. 
For this we need to exclude the $U$'s for which $D^U$ is ill-defined. Let $\U$ denote the space of 
lattice gauge fields resulting from excluding this measure zero subspace from $\U_{\rm initial}$.
A sufficient condition for $U$ to be in $\U$
is that the plaquette variables satisfy $||1-U(p)||<\epsilon$
for some sufficiently small $\epsilon$ \cite{L}. Since $1-U(p)=a^2F_{\mu\nu}(x)+O(a^3)$ the lattice 
transcript of any smooth continuum field $A\in\A$ is guaranteed to lie in $\U$ when the lattice is 
sufficiently fine. The lattice $Q$ reduces to the continuum $Q$ in the classical continuum limit
\cite{DA(JMP)}. $D^U$ and $Q$ have their origins in the overlap formalism \cite{ov}.

Let $\C$ denote the space of lattice spinor fields on $T^4$; for simplicity we assume that the fermion
is in the fundamental representation of $\SU(N)$. $\C$ has two ``chiral'' decompositions:
\be
\C=\C_+\oplus\C_-\qquad\mbox{and}\qquad\C=\hC_+^U\oplus\hC_-^U 
\nonumber
\ee
defined, respectively, by $\g5=\pm1$ on $\C_{\pm}$ and $\hg5^U=\pm1$ on $\hC_{\pm}^U$ where 
$\hg5^U=-\mbox{sign}(H^U)\,$ ($H^U$=the Hermitian Wilson-Dirac operator with suitable negative mass
term). It can be shown that \cite{ov}
\be
\index\,D^U=\mbox{dim}\,\hC_+^U-\mbox{dim}\,\C_-
\label{10}
\ee
This leads to a natural definition of the index bundle of the overlap Dirac operator \cite{DA(NPB1)}:
\be
\index\,D=\hC_+-\C_-
\label{11}
\ee
$\hC_+=\{\hC_+^U\}_{U\in\U}$ is a vector bundle over $\U$ \cite{DA(NPB1)} and in (\ref{11}) $\C_-$
is the trivial bundle over $\U$ with constant fiber $\C_-$. Due to gauge covariance of $D^U$
the bundle $\hC_+\,$, and therefore also $\index\,D$, descend to bundles over the orbit space $\U/\G$.

Consider now a family $U^{(y)}$ ($y\in{}Y$) in $\U$ with the property that
the fields parameterized by $y\in\partial{}Y$
all gauge equivalent. It determines a closed submanifold in the orbit space $\U/\G$ and the topological
charge of the family can be defined fermionically in analogy with (\ref{3}):
\be
Q_Y:=\int_{Y/\partial{}Y}\,\mbox{ch}(\index\,D)
\nonumber
\ee
The topological properties of the Chern character guarantee that this is an integer.
We now specialize to $Y=B^{2p}\,$, $\partial{}Y=S^{2p-1}$ so that $U^{(y)}\equiv{}U^{(\theta,t)}$
with $U^{(\theta,1)}\!=\!\phi_{\theta}\cdot{}U$. An explicit formula for the families topological charge 
in this case has been derived in \cite{DA(NPB1)}: For $p>0$,
\be
Q_{2p}={\textstyle \frac{1}{(2\pi{}i)^p}}
\Big({\textstyle \frac{1}{p!}}\int_{B^{2p}}
\Tr\left\lb{}P^{(\theta,t)}(dP^{(\theta,t)})^{2p}\right\rb\qquad \nonumber \\ 
\quad\ \ +\ {\textstyle \frac{(-1)^p}{2}\frac{(p-1)!}{(2p-1)!}}
\int_{S^{2p-1}}\Tr\left\lb\hg5^U
(\phi_{\theta}^{-1}d_{\theta}\phi_{\theta})^{2p-1}\right\rb\Big) 
\nonumber \\
\qquad+\ 2\sum_xdeg(\phi(x))\qquad\qquad\qquad\qquad\qquad
\nonumber
\ee
Here $\hg5=\hg5^{U^{(\theta,t)}}\,$, $P=\frac{1}{2}(1+\hg5)$, $d\!=\!d_{\theta}\!+\!d_t$ is 
the exterior derivative on $B^{2p}\,$, $\Tr$ is the trace for linear operators on $\C$, and
$deg(\phi(x))$ is the degree of the map $S^{2p-1}\to\mbox{SU(N)}\,$, $\,\theta\mapsto\phi_{\theta}(x)$.
From this formula one can show the following \cite{DA(NPB1)}:

\vspace{1ex}

\noindent {\bf Theorem}. The lattice $Q_{2p}$ reduces to the continuum $Q_{2p}$ in the classical 
continuum limit.

\vspace{1ex}

\noindent In light of the remarks below Eq.(\ref{7}), examples of families $U^{(y)}$ in $\U$
can be obtained as the lattice transcripts of families $A^{(y)}$ in $\A$ when the lattice is sufficiently 
fine. In particular, specific examples of $B^{2p}$-families are obtained as the lattice transcripts
of (\ref{4}). It follows from the Theorem that when $Q_{2p}$ for the continuum family is nonvanishing
then the lattice $Q_{2p}$ is also nonvanishing for the transcripted family, at least when the lattice is 
sufficiently fine. So lattice families with nonvanishing $Q_{2p}$ do exist.

{\it Physical significance of the lattice $Q_{2p}$'s}. In complete analogy with the continuum situation
$Q_2$ is an obstruction to gauge invariance of the overlap chiral fermion determinant
\cite{DA(NPB1)}. However, in contrast to the continuum 
situation, the lattice $Q_{2p}$'s for general $p>0$ are not obstructions to the existence of gauge
fixings without the Gribov problem. Such gauge fixings do exist on the lattice; examples of these are
the maximal tree gauges introduced in \cite{Creutz}. Hence a decomposition 
\be
\U\simeq\U/\G\times\G
\label{14}
\ee
does exist on the lattice. (To make the action of $\G$ on $\U$ free we are again imposing the condition
$\phi(x_0)\!=\!1$ on the gauge transformations.) 
The difference between the continuum and lattice situations is due to the fact 
that, while all spheres in $\A$ are contractible, the same is not true for $\U$. Indeed, by turning around 
our previous continuum argument we see that, on the lattice, nonvanishing $Q_{2p}$ implies 
noncontractibility of the $2p$-sphere ${\cal S}^{2p}$ in $\U/\G$ determined by the $B^{2p}$-family 
$U^{(\theta,t)}$. This in turn implies, via (\ref{14}), the noncontractibility of a ``gauge fixed''
$2p$-sphere ${\cal S}_f^{2p}$ in $\U$ arising as the image in $\U$ of 
${\cal S}^{2p}\times\{1\}$ in $\U/\G\times\G$. 

Thus we see that the obstructions $Q_{2p}$ to gauge fixing 
without the Gribov problem in the continuum correspond on the lattice to obstructions to the 
contractibility of certain $2p$-spheres in $\U$. It follows from this and the Theorem above that
\be
\pi_{2p}(\U)\ne0\qquad\quad\mbox{for}\ 1\le{}p\le{}N-2
\nonumber
\ee
at least when the lattice is sufficiently fine. This is a direct consequence of excluding the lattice
gauge fields for which the fermionic topological charge $Q\!=\!\index\,D^U$ is ill-defined. To see this
note that, by (\ref{6a}), $\pi_{2p}(\U_{\rm initial})\!=\!0$ for 
$1\!\le\!{}p\!\le{}\!N\!-\!1$ since the same is true
for $\pi_{2p}(\SU(N))$. Hence the noncontractible $2p$-spheres in $\U$ are all contractible in
$\U_{\rm initial}$. For more on all this, including an explicit description of the noncontractible
$2p$-spheres ${\cal S}_f^{2p}$ in $\U$, see \cite{DA(NPB2)}.

In \cite{DA(NPB2)} it was shown that $Q_{2p}$ coincides with the families topological
charge $(Q_{2p})_f:=\int_{{\cal S}_f^{2p}}\,\mbox{ch}(\index\,D)$ associated with the 
$2p$-sphere ${\cal S}_f^{2p}$ in $\U$. In the $2p\!=\!2$ case this implies that the 
obstructions $Q_2$ to gauge invariance of the overlap chiral fermion determinant coincide with 
obstructions $(Q_2)_f$ to trivialising the overlap determinant line bundle over 2-sphere ${\cal S}_f^2$
in $\U$. The situation is the unchanged for more general 2-dimensional families $U^{(y)}$ in $\U\,$:
the closed 2-manifold ${\cal S}$ in $\U/\G$ determined by the family is isomorphic via (\ref{14})
to a ``gauge-fixed'' 2-manifold ${\cal S}_f$ in $\U$ itself, and the families topological charges
$Q_Y$ and $(Q_Y)_f$ coincide. A recent observation in \cite{KS} can be understood in this context.
An obstruction to trivialising the overlap determinant line bundle over a certain torus in $\U$
was described there; it coincides with the obstruction to trivialising the overlap over a torus in
$\U/\G$ discussed previously by Neuberger \cite{Neu(geom)} and this is an example of the situation that
we have just discussed.

\end{document}